\documentclass[pra,twocolumn,showpacs,amsmath,amssymb]{revtex4}
\usepackage{epsfig}
\usepackage{color}
\begin{document}

\title{Phase diagram of spin-$1/2$ bosons in one-dimensional optical
  lattice}
\author{L. de Forges de Parny$^1$, M. Traynard$^1$, F. H\'ebert$^1$,
  V.G. Rousseau$^2$, R.T. Scalettar$^3$, and G.G. Batrouni$^{1,4}$}  
\affiliation{
$^1$INLN, Universit\'e de Nice-Sophia Antipolis, CNRS; 
1361 route des Lucioles, 06560 Valbonne, France
}
\affiliation{
$^2$Department of Physics and Astronomy, Louisiana State University,
  Baton Rouge, Louisiana 70803, USA 
}
\affiliation{
$^3$Physics Department, University of California, Davis, CA 95616
}
\affiliation{
$^4$Centre for Quantum Technologies, National University of Singapore; 2
Science Drive 3 Singapore 117542.
}

\begin{abstract}
Systems of two coupled bosonic species are studied using Mean Field
Theory and Quantum Monte Carlo.  The phase diagram is characterized
both based on the mobility of the particles (Mott insulating or
superfluid) and whether or not the system is magnetic (different
populations for the two species).  The phase diagram is shown
to be population balanced for negative spin-dependent interactions,
regardless of whether it is insulating or superfluid.  For positive
spin-dependent interactions, the superfluid phase is always polarized,
the two populations are imbalanced. On the other hand, the
Mott insulating phase with even commensurate filling has balanced
populations while the odd commensurate filling Mott phase has balanced
populations at very strong interaction and polarizes as the interaction
gets weaker while still in the Mott phase.
\end{abstract}


\maketitle

\section{Introduction}

When ultra-cold bosonic atoms with internal degrees of freedom, {\it
e.g.~}$^{87}$Rb or $^{23}$Na in $F=1$ hyperfine state, are confined
in magneto-optical traps, the magnetic field aligns the spins and the
system behaves effectively as one of spin-$0$ particles. Progress in
purely optical trapping techniques now allows the confinement of these
atoms with their full spin degrees of freedom present
\cite{stamper1,stamper2} which permits the study of the interplay
between magnetism and superfluidity. In such systems, in addition to
the usual spin-independent contact interaction, there are
spin-dependent terms \cite{ho} and also longer range interactions
\cite{stamper3}. Mean field calculations \cite{ho} in the absence of
the long range interactions, show that the interactions can be
ferromagnetic ({\it e.g.~}$^{87}$Rb) or antiferromagnetic 
({\it e.g.~}$^{23}$Na), depending on the relative magnitudes of the
scattering
lengths in the singlet and quintuplet channels. When loaded in an
optical lattice, the system is governed by an extended Hubbard
Hamiltonian with spin-dependent and spin-independent terms. The phase
diagram in the mean field and variational approximations has been
calculated at zero 
\cite{imambekov04,tsuchiya05,zhou,pai08} and at finite temperature
\cite{pai08}.

The overall picture emerging from these calculations is that in the
ferromagnetic case, {\it i.e.} negative spin-dependent interaction,
the system is in a Mott insulating (MI) phase when the filling is
commensurate with the lattice size and the repulsive
contact interaction is strong enough. When this interaction is weaker,
or when the particle filling is incommensurate, the
system is in a ferromagnetic superfluid (SF) phase. The SF-MI
transitions in the ground state are all expected to be
continuous. Furthermore, all Mott lobes shrink, and eventually
disappear, as the ratio of the spin-dependent interaction to the
spin-independent one increases. These predictions have been confirmed
for the one-dimensional case with Quantum Monte Carlo (QMC)
simulations \cite{batrouni}.

The phase diagram for the antiferromagnetic case also has MI lobes at
commensurate fillings and
strong coupling and a SF phase at weak coupling or incommensurate
fillings. However, the nature of the SF phase and the MI-SF
transitions are different. In this case, the SF phase is ``polar'':
Superfluidity is carried either by the $S_z=0$ component or by the
$S_z=\pm 1$ components. Furthermore, the Mott lobes of odd order are
expected to be nematic in two and three dimensions and dimerized in
one dimension \cite{imambekov04}. The MI-SF transition into these
odd Mott lobes is generally expected to be continuous. In the even
Mott lobes, the bosons are predicted to be in a singlet state which is
expected to increase the stability of the MI. This transition from
non-singlet SF to singlet MI is argued to render the transition into
the even Mott lobes of first order. Density Matrix Renormalization
Group (DMRG) \cite{dmrg1,dmrg2} and QMC \cite{apaja06} calculations
for one-dimensional systems have confirmed the dimerized nature of
odd lobes, and determined the phase diagram. Further QMC
work \cite{batrouni} presented evidence of singlet formation in the
even lobes but did not find clear evidence of first order phase
transitions.

A related model is obtained by coupling degenerate atomic ground
states and degenerate excited states (see next section) to yield a
system of two bosonic species sometimes referred to as spin-$1/2$
bosons \cite{krutitsky04,ashhab}. As for the full spin-$1$ case,
the spin-dependent interaction can be ferromagnetic or
antiferromagnetic. Mean field analysis of this model \cite{krutitsky04}
yielded phase diagrams which are rather similar to the spin-$1$ case. In
this paper we shall present an extension of previous
mean field treatments 
\cite{krutitsky04} which yields better
qualitative agreement with exact QMC simulations, which we also present
for the one-dimensional system. The paper is organized as follows:
In Sec.~II the Hamiltonian of two bosonic species is presented.
The Mean Field phase diagram is discussed in Sec.~III. 
Finally, in Sec.~IV Quantum Monte Carlo calculations on one dimensional
lattices verify the qualitative conclusions of the 
Mean Field theory (MFT), but provide quantitatively
accurate values for the phase boundaries.

\section{Two Bosonic Species Hamiltonian}

We follow Ref.~\cite{krutitsky04} and consider a system of neutral
polarizable bosonic atoms with three degenerate internal ground states
($F_g=1$) and excited states ($F_e=1$) characterized by the magnetic
quantum number $S_z=0,\pm 1$. The atoms are loaded in a
$d$-dimensional optical lattice produced by counterpropagating laser
beams. As explained in Ref.~\cite{krutitsky04}, in addition to
generating the optical lattice, the lasers can be used to couple the
internal ground and excited states by $V$ and $\Lambda$ transitions
leading to two sets of orthogonal Bloch eigenmodes denoted
respectively by $0$ and $\Lambda$ \cite{krutitsky04}. The resulting
Bose Hubbard Hamiltonian is given by
\begin{eqnarray}
\nonumber
H&=&-t\sum_{\langle i,j\rangle,\sigma} \left (a^\dagger_{\sigma i}
a^{\phantom\dagger}_{\sigma j} + a^\dagger_{\sigma j}
a^{\phantom\dagger}_{\sigma i}\right ) + \frac{U_0}{2}\sum_{\sigma,i}
{\hat n}_{\sigma i} \left ( {\hat n}_{\sigma i}-1\right )\\
\nonumber
&&+ \frac{U_2 }{2}{\rm cos}(\delta\phi) \sum_i
\left(a^\dagger_{0i}a^\dagger_{0i}a^{\phantom\dagger}_{\Lambda
  i}a^{\phantom\dagger}_{\Lambda i} + a^\dagger_{\Lambda
  i}a^\dagger_{\Lambda i}a^{\phantom\dagger}_{0 i}a^{\phantom\dagger}_{0 i}
\right )\\
&& + (U_0+U_2)\sum_i {\hat n}_{0i}{\hat n}_{\Lambda i} -\mu
\sum_{\sigma,i}{\hat n}_{\sigma i},
\label{ham1}
\end{eqnarray}
where $a^\dagger_{\sigma i}$ ($a^{\phantom\dagger}_{\sigma i}$)
creates (annihilates) a particle of ``spin'' $\sigma=0,\Lambda$ on site $i$,
$U_0>0$ is the spin-independent contact repulsion term and $U_2$ is
the spin-dependent interaction term. The number operator, ${\hat
  n}_{\sigma i}$, counts the number of particles of type $\sigma$ on
site $i$ and $\mu$ is the chemical potential. We take the hopping
parameters $t=1$ to set the energy scale. The phase difference,
$\delta\phi =2(\phi_0-\phi_{\Lambda})$, gives the relative global
phase between the two species. Equation (\ref{ham1}) is the same as
Eq.~(9) of Ref.~\cite{krutitsky04} after one uses Eqs.~(13) and (15) of
that reference. The relative phase, $(\phi_0-\phi_\Lambda)$, was then
determined by the requirement that the total energy be minimized. It
was argued in Ref.~\cite{krutitsky04} that since $U_2$ can be positive
or negative, the way to minimize the energy is to have the coupling in
the third term of Eq.~(\ref{ham1}) be of the form $-|U_2|/2$. This
gives $\phi_0=\phi_{\Lambda}$ for $U_2<0$ and
$\phi_0=\phi_{\Lambda}\pm \pi/2$ for $U_2>0$. The Hamiltonian then
becomes
\begin{eqnarray}
\nonumber
H_1&=&-t\sum_{\langle i,j\rangle,\sigma} \left (a^\dagger_{\sigma i}
a^{\phantom\dagger}_{\sigma j} + a^\dagger_{\sigma j}
a^{\phantom\dagger}_{\sigma i}\right ) + \frac{U_0}{2}\sum_{\sigma,i}
{\hat n}_{\sigma i} \left ( {\hat n}_{\sigma i}-1\right )\\
\nonumber
&&- \frac{|U_2|}{2}\sum_i
\left(a^\dagger_{0i}a^\dagger_{0i}a^{\phantom\dagger}_{\Lambda 
  i}a^{\phantom\dagger}_{\Lambda i} + a^\dagger_{\Lambda
  i}a^\dagger_{\Lambda i}a^{\phantom\dagger}_{0 i}a^{\phantom\dagger}_{0 i}
\right )\\
&& + (U_0+U_2)\sum_i {\hat n}_{0i}{\hat n}_{\Lambda i} -\mu
\sum_{\sigma,i}{\hat n}_{\sigma i}.
\label{ham2}
\end{eqnarray}
However, this is not the only possible solution: The sign of the third
term has no effect on the energy and the Hamiltonian obtained by
putting $\phi_0=\phi_{\Lambda}$ both for $U_2$ positive or negative,
\begin{eqnarray}
\nonumber
H_2&=&-t\sum_{\langle i,j\rangle,\sigma} \left (a^\dagger_{\sigma i}
a^{\phantom\dagger}_{\sigma j} + a^\dagger_{\sigma j}
a^{\phantom\dagger}_{\sigma i}\right ) + \frac{U_0}{2}\sum_{\sigma,i}
{\hat n}_{\sigma i} \left ( {\hat n}_{\sigma i}-1\right )\\
\nonumber
&&+ \frac{U_2}{2}\sum_i
\left(a^\dagger_{0i}a^\dagger_{0i}a^{\phantom\dagger}_{\Lambda 
  i}a^{\phantom\dagger}_{\Lambda i} + a^\dagger_{\Lambda
  i}a^\dagger_{\Lambda i}a^{\phantom\dagger}_{0 i}a^{\phantom\dagger}_{0 i}
\right )\\
&& + (U_0+U_2)\sum_i {\hat n}_{0i}{\hat n}_{\Lambda i} -\mu
\sum_{\sigma,i}{\hat n}_{\sigma i},
\label{ham3}
\end{eqnarray}
yields exactly the same free energy as Eq.~(\ref{ham2}). For the
$U_2<0$ case, this is obvious since Eq.~(\ref{ham2}) is identical to
Eq.~(\ref{ham3}). For $U_2>0$, this can be seen by writing the
partition function, $Z= {\rm Tr}\,{\rm e}^{-\beta H}$, and noting that
if we expand the exponential in powers of the term in question, only
even powers will contribute. In fact, for $U_2>0$, one can transform
Eq.~(\ref{ham2}) into Eq.~(\ref{ham3}) by a simple phase transformation,
$a_{0i}\to a_{0 i}{\rm e}^{i\pi/2}$.  Therefore, both Eqs.~(\ref{ham2})
and (\ref{ham3}) are valid and yield the same results for
energies and all other quantities that are invariant under the 
above phase transformation when examined
with an exact method such as QMC. This is not necessarily so when
approximations are made as we shall see in the next section.

\section{Mean Field Phase Diagrams}

In this section we revisit the mean field phase diagrams of the model
governed by Eq.~(\ref{ham2}) and (\ref{ham3}). This is implemented by
first writing the identity,
\begin{eqnarray}
\nonumber
a^{\dagger}_{\sigma i}a^{\phantom\dagger}_{\sigma j} &=&
  \left (a^{\dagger}_{\sigma i}-\langle a^{\dagger}_{\sigma}
  \rangle\right )
  \left (a^{\phantom\dagger}_{\sigma j}-\langle
  a^{\phantom\dagger}_{\sigma} \rangle \right ) \\
&&+ \langle
  a^{\dagger}_{\sigma} \rangle a^{\phantom\dagger}_{\sigma j} +
  a^{\dagger}_{\sigma i} \langle
  a^{\phantom\dagger}_{\sigma} \rangle - \langle
  a^{\dagger}_{\sigma}\rangle \langle a^{\phantom\dagger}_{\sigma}
  \rangle.
\label{mfapprox}
\end{eqnarray}
The superfluid order parameter is $\langle a^{\dagger}_{\sigma}
\rangle = \langle a^{\phantom\dagger}_{\sigma} \rangle \equiv
\psi_{\sigma}$ and is taken to be real and uniform, {\it i.e.}
independent of the position.  The first term in Eq.~(\ref{mfapprox}) is
a small fluctuation and is ignored;
substituting the remaining terms in Eq.~\ref{ham2} reduces $H_1$ to the
sum of single site Hamiltonians,
\begin{eqnarray}
\nonumber
H_{MF1}&=& -2td \sum_\sigma \left( \psi_\sigma  (a^{\dagger}_{\sigma} +
a^{\phantom\dagger}_{\sigma}) - \psi_\sigma^2 \right ) \\
\nonumber
&& + \frac{U_0}{2}\sum_{\sigma}
{\hat n}_{\sigma} \left ( {\hat n}_{\sigma}-1\right )\\
\nonumber
&&- \frac{|U_2|}{2}
\left(a^\dagger_{0}a^\dagger_{0}a^{\phantom\dagger}_{\Lambda} 
a^{\phantom\dagger}_{\Lambda} + a^\dagger_{\Lambda}a^\dagger_{\Lambda}
a^{\phantom\dagger}_{0}a^{\phantom\dagger}_{0}
\right )\\
&& + (U_0+U_2){\hat n}_{0}{\hat n}_{\Lambda} -\mu
\sum_{\sigma}{\hat n}_{\sigma};
\label{mftham2}
\end{eqnarray}
and from Eq.~(\ref{ham3}), $H_2$ becomes
\begin{eqnarray}
\nonumber
H_{MF2}&=& -2td \sum_\sigma \left( \psi_\sigma (a^{\dagger}_{\sigma} +
a^{\phantom\dagger}_{\sigma}) - \psi_\sigma^2 \right ) \\
\nonumber
&& + \frac{U_0}{2}\sum_{\sigma}
{\hat n}_{\sigma} \left ( {\hat n}_{\sigma}-1\right )\\
\nonumber
&&+ \frac{U_2}{2}
\left(a^\dagger_{0}a^\dagger_{0}a^{\phantom\dagger}_{\Lambda} 
a^{\phantom\dagger}_{\Lambda} + a^\dagger_{\Lambda}a^\dagger_{\Lambda}
a^{\phantom\dagger}_{0}a^{\phantom\dagger}_{0}
\right )\\
&& + (U_0+U_2){\hat n}_{0}{\hat n}_{\Lambda} -\mu
\sum_{\sigma}{\hat n}_{\sigma}.
\label{mftham3}
\end{eqnarray}
The mean field solution is obtained by taking matrix elements of these
Hamiltonians in a truncated basis of single site occupation number
states $| \, n_0 \, n_{\Lambda} \, \rangle$, 
diagonalizing the resulting matrix and then minimizing
the lowest eigenvalue with respect to the order parameters
$\psi_{\sigma}$. The basis is truncated for large particles number
(typically $n_0 = n_\Lambda = 12$)
to obtain results independent of the truncation.  
This gives the order parameters of the ground state
and its eigenvector. The superfluid density is given by
\begin{equation}
 \rho_s = \psi_0^2 + \psi_{\Lambda}^2.
\label{rhosMFT}
\end{equation}

\subsection{Ferromagnetic: $U_2<0$}

\begin{figure}[ht]
\epsfig{figure=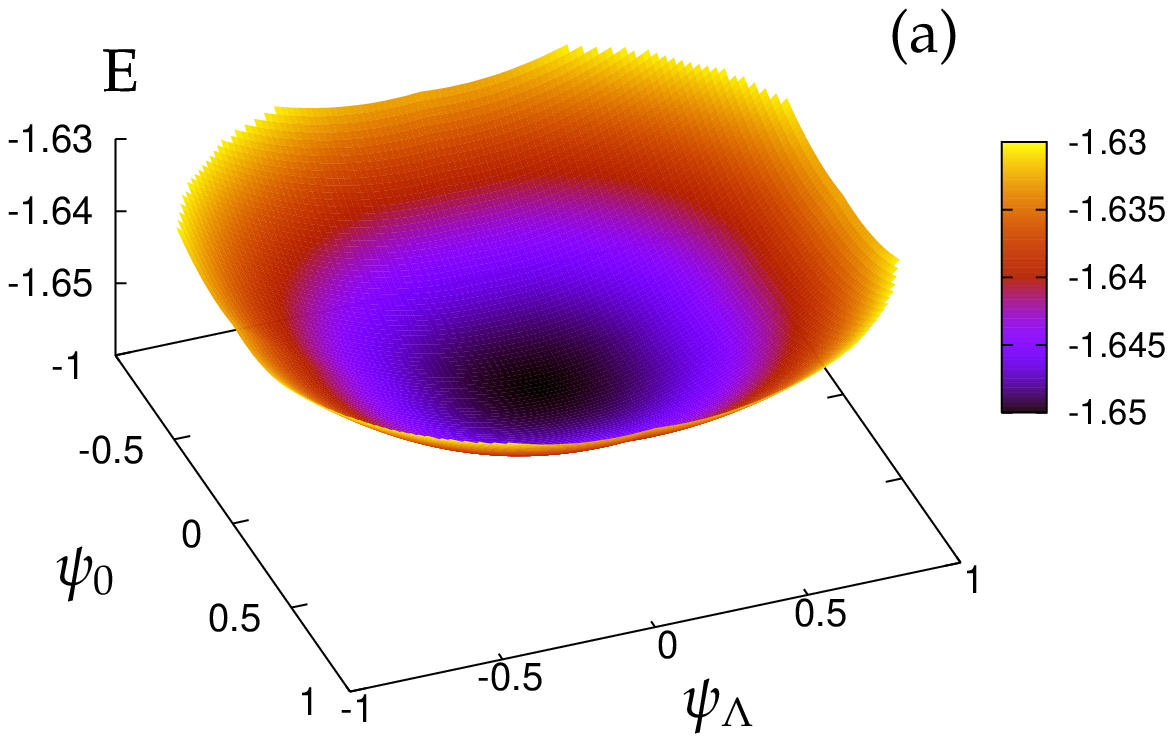,width=5.5cm,
clip }
\epsfig{figure=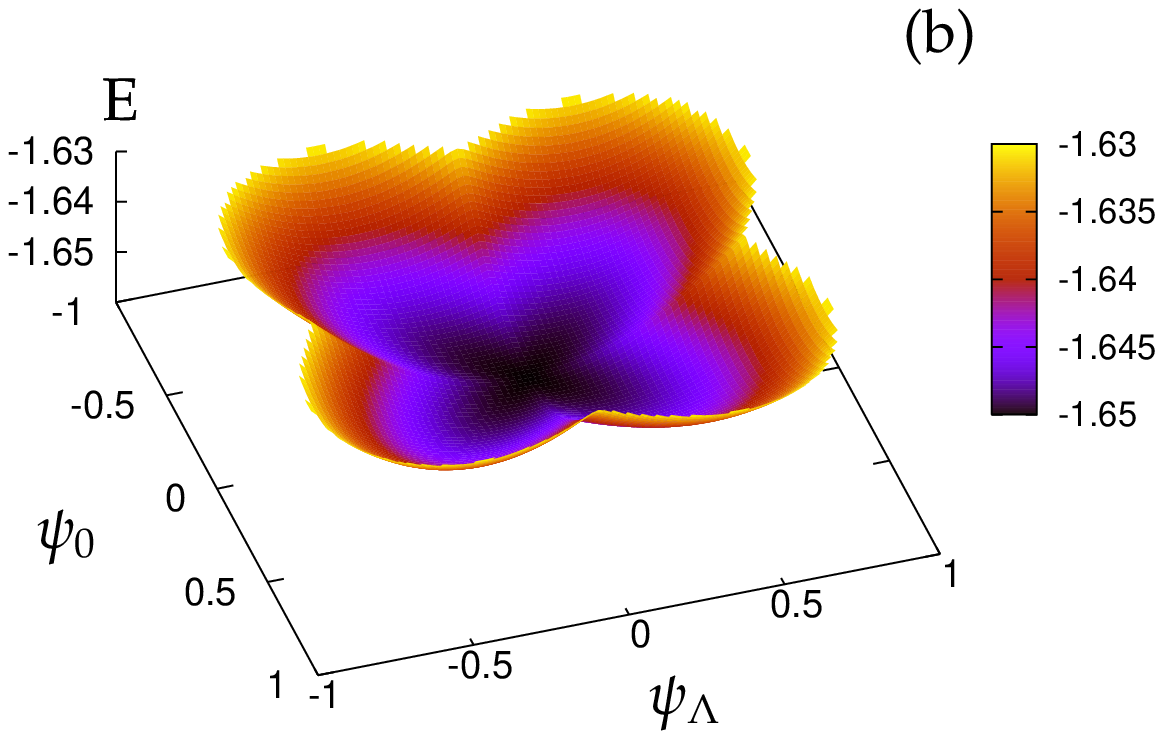,width=5.5cm,clip
}
\epsfig{figure=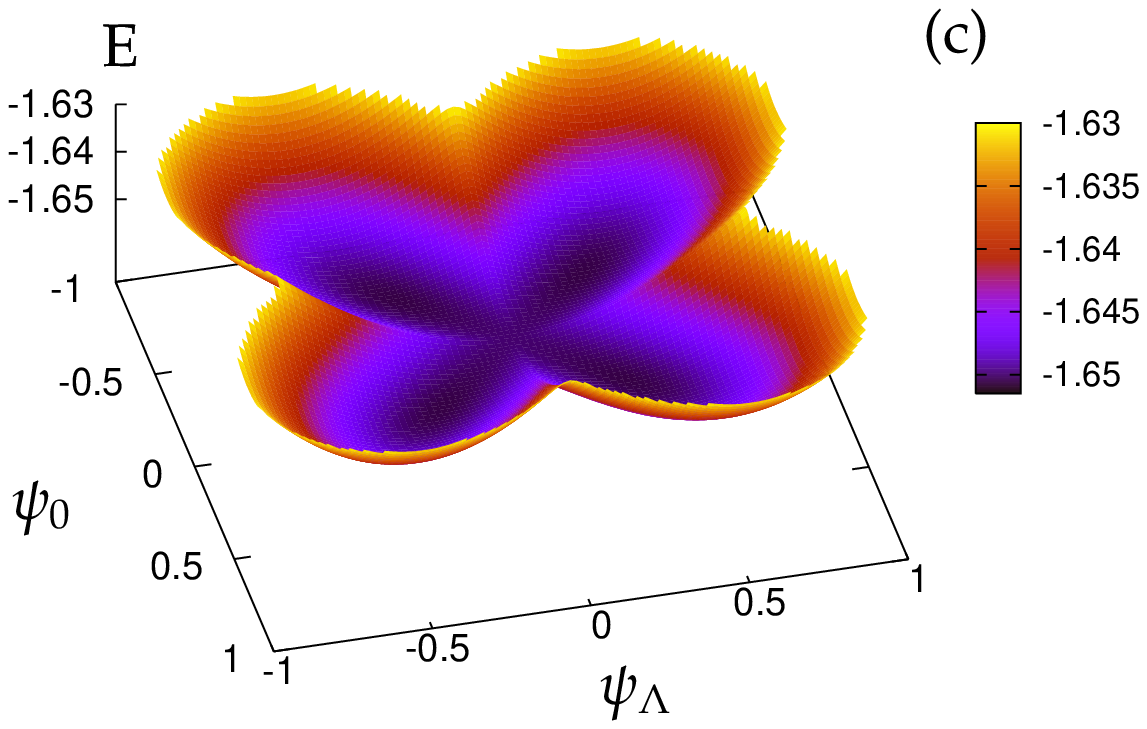,width=5.5cm,clip
}
\epsfig{figure=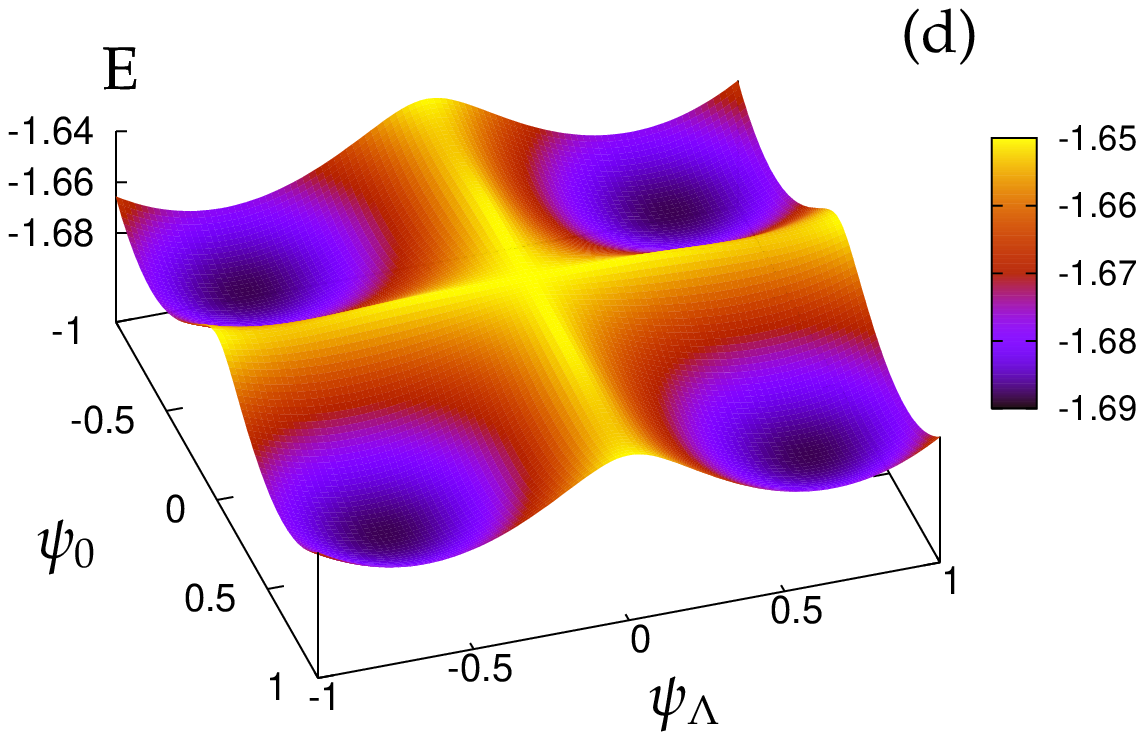,width=5.5cm,
clip }
\caption{(Color online) The ground state energy, $E$, given
by Eq.~(\ref{mftham2}) or (\ref{mftham3}) in the ferromagnetic case,
$U_2=-0.1U_0$ at $\rho=2$ ($\mu/U_0=1.3$). (a) $U_0=100t$ the system
is in the MI phase. (b) $U_0=25t$, the system is in the MI phase close
to the transition into the SF phase. (c) $U_0=20t$, the system has just
made the transition into the SF phase. The minimum has changed
continuously from
$\psi_0=\psi_{\Lambda}=0$ to four degenerate minima at nonzero
values of the order parameters. (d) $U_0=12.5t$, the system is in
the superfluid phase. Note the degenerate symmetric minima in the SF
case, $|\psi_0|=|\psi_{\Lambda}|\neq 0$. This figure agrees with Fig. 2 in
Ref.~\cite{krutitsky04}.}
\label{ferroEmin}
\end{figure}

The two mean field Hamiltonians, $H_{MF1}$ and $H_{MF2}$ are identical
for the ferromagnetic case, $U_2<0$, and their ground state was studied in Ref.
\cite{krutitsky04}. At commensurate filling and strong interaction (small
$t/U_0$) the energy has one minimum at $\psi_0=\psi_{\Lambda}=0$ corresponding
to the MI. Figure \ref{ferroEmin}(a) shows this for the Mott phase with
two particles/site, $\rho=2$. For any non commensurate filling and
for commensurate filling at weak coupling, there are four symmetric and
degenerate minima with $|\psi_0|=|\psi_{\lambda}|\neq 0$ corresponding to
the superfluid phase, Fig. \ref{ferroEmin}(d). For later reference, we
emphasize that the superfluid phase is symmetric in the two particle species,
$|\psi_0|=|\psi_{\lambda}|$. Figures \ref{ferroEmin}(b,c) show how the
minimum at $\psi_0=\psi_{\Lambda}=0$ transforms continuously into
four degenerate minima visible in Fig. \ref{ferroEmin}(d) as the
interaction gets weaker. The same behaviour is seen for the MI at all
other commensurate fillings.

\begin{figure}[ht]
\epsfig{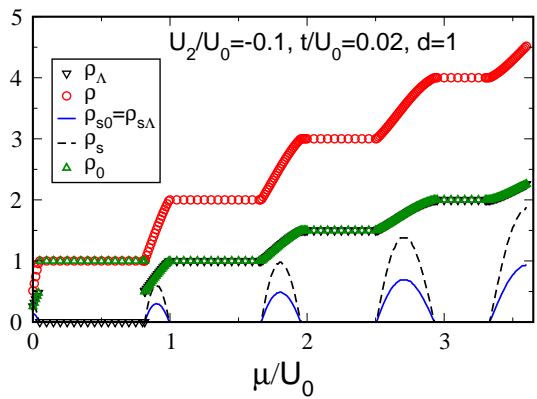}
\caption{(Color online) The densities of the two species, $\rho_0$ and
$\rho_{\Lambda}$, the total density, $\rho=\rho_0+\rho_{\Lambda}$, and
the corresponding superfluid densities as functions of $\mu/U_0$. In
the ferromagnetic case, $\rho_{s0}=\rho_{s\Lambda}$ and
$\rho_0=\rho_{\Lambda}$ except in the first Mott lobe. The reason is
that the particles do not interact and so there is no possibility to
convert one species to another.}
\label{ferrorhovsmuMFT}
\end{figure}

\begin{figure}[ht]
\epsfig{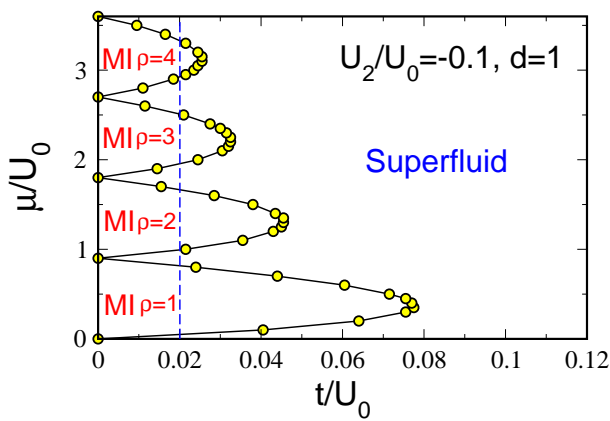}
\caption{(Color online) The phase diagram given by the mean field
Hamiltonians, Eq.~(\ref{mftham2}) or (\ref{mftham3}), in the
ferromagnetic case $U_2=-0.1U_0$. Note the shrinking of the bases of
all lobes by $|U_2|/U_0$. The dashed vertical line shows where the
cut in Fig.~\ref{ferrorhovsmuMFT} was taken.}
\label{ferrophasediagMFT}
\end{figure}

The MF phase diagram is obtained by repeating the above calculation
for many values of the chemical potential, $\mu$, and the
interaction $U_0$ (always keeping $U_2/U_0$ constant). In the
no-hopping limit, $t/U_0 \to 0$, it is easy to see that when
$\mu$ satisfies the condition,
\begin{equation}
(n-1)\left (U_0-|U_2|\right ) <\mu <n\left (U_0-|U_2|\right ),
\label{mottbaseferro}
\end{equation}
the system is in a MI phase with $n$ bosons per site. As $t/U_0$
increases, always keeping $|U_2|/U_0$ constant, the MI region shrinks
and eventually disappears giving the familiar Mott lobes. Outside the
MI the system is superfluid. Note that $U_2$ has the effect of
shrinking the bases of all the Mott lobes,
Fig. \ref{ferrophasediagMFT}: When $|U_2|=U_0$ the lobes disappear
completely. In Fig. \ref{ferrorhovsmuMFT} we show mean field results
for the densities, $\rho_0$ and $\rho_{\Lambda}$ of the two species,
the total density, $\rho=\rho_0+\rho_{\Lambda}$ and the
corresponding superfluid densities as functions of $\mu$ for
fixed $t/U_0=0.02$ and $U_2/U_0=-0.1$. We see that all compressible
regions, $\kappa=\partial \rho/\partial \mu \neq 0$, are superfluid
while the incompressible plateaux, $\kappa=0$, are not superfluid,
they are the MI. The absence of discontinuous jumps in any of
the densities as the system enters the MI indicates the transitions
are continuous as also discussed in Ref.~\cite{krutitsky04}. We
also remark that since in the ferromagnetic case we
have $|\psi_0|=|\psi_{\Lambda}|$, then $\rho_{s0}=\rho_{s\Lambda}$
and $\rho_0=\rho_{\Lambda}$ everywhere except in the first Mott
plateau where one of the species has zero density, in this
case $\rho_{\Lambda}=0$. The reason the first Mott region behaves 
this way is that in order for the two species to interconvert, the
particles have to meet and interact. In the MI phase, the particles do
not hop between the sites with the consequence that in the first MI
there is only one particle per site and no way to interconvert the
species. The phase diagram is shown in Fig.~\ref{ferrophasediagMFT}.

\subsection{Anti-ferromagnetic: $U_2>0$}

In the anti-ferromagnetic case, $U_2>0$, the Hamiltonians
Eq.~(\ref{mftham2}) and Eq.~(\ref{mftham3}) are no longer
identical. Reference \cite{krutitsky04} performed the mean
field calculation using Eq.~(\ref{mftham2}) and found that in the MI, the
energy has one minimum at $\psi_0=\psi_{\Lambda}=0$. As the system
is taken closer to the tip of the Mott lobe, the energy develops
local minima (Fig.~6, Ref.~\cite{krutitsky04}); these local minima become
global minima in the SF phase indicating that the MI-SF transition can
be of first order.  Furthermore, it was found that there is a continuum
of degenerate minima at nonzero $\psi_0$ and $\psi_{\Lambda}$ (Fig.~6, 
Ref.~\cite{krutitsky04}).

Doing the mean field calculation with Eq.~(\ref{mftham3}) leads to
results which are similar in some ways to the above but different in
very important aspects.  First, for a given choice of $t,\,U_0,\,U_2$
and $\mu$, we found that the two mean field Hamiltonians give the
same ground state energy both in the SF and MI phases. Therefore, one
cannot choose between them on this basis. In the MI phase, whether the
filling is odd or even, Eq.~(\ref{mftham3}) gives a ground state energy
with one minimum at $\psi_0=\psi_{\Lambda}=0$; the same as
Eq.~(\ref{mftham2}). As the system is brought closer to the lobe tip,
the behavior depends on whether the filling in the Mott phase is odd or
even.  Odd filling is similar to the ferromagnetic
case: The global minimum at $\psi_0=\psi_{\Lambda}=0$ transforms
continuously into four degenerate minima with non-vanishing
order parameter. However, in this anti-ferromagnetic case, the minima
lie on the $\psi_0$ and $\psi_{\Lambda}$ axes: Either $\psi_0=0$ while
$\psi_{\Lambda}\neq 0$, or $\psi_0\neq 0$ while $\psi_{\Lambda}= 0$.
This means that in the SF phase, the SF density is
composed entirely of one species: Either
$(\rho_s=\rho_{s0},\,\rho_{s\Lambda}=0)$ or
$(\rho_{s0}=0,\,\rho_s=\rho_{s\Lambda})$. Our result differs from that
in Ref.~\cite{krutitsky04} where a continuum of degenerate minima was
found in the SF phase.

\begin{figure}[ht]
\epsfig{figure=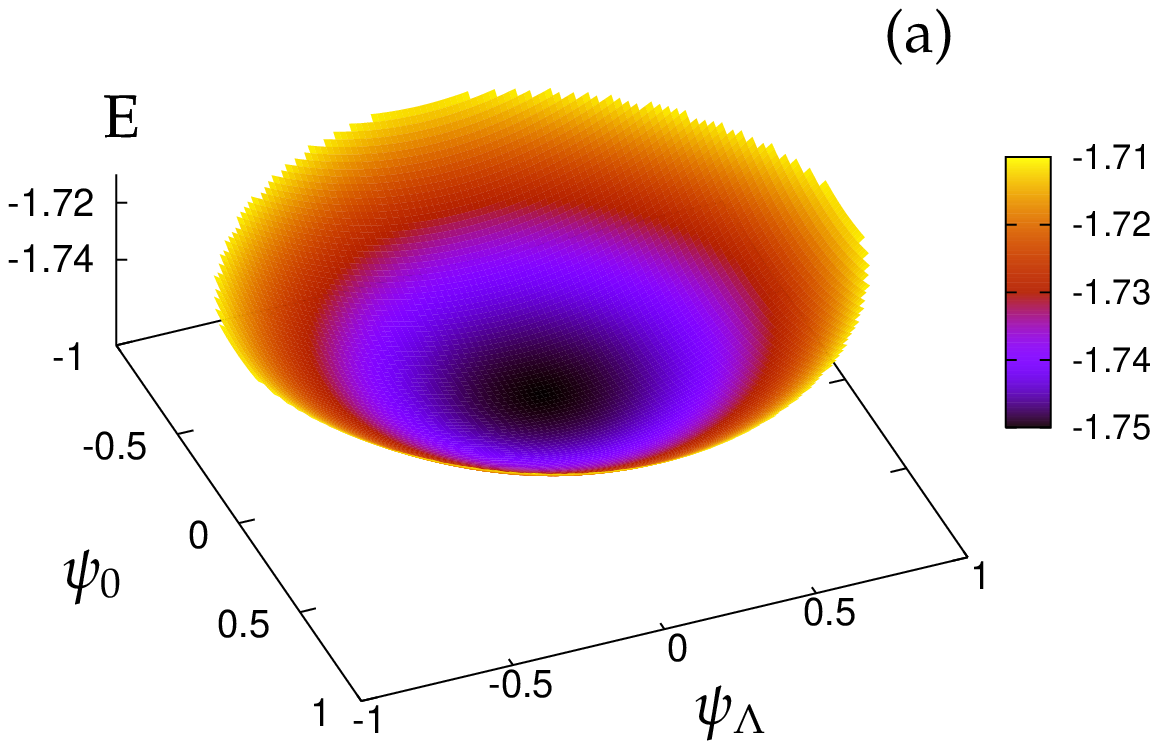,width=6cm,
clip}
\epsfig{figure=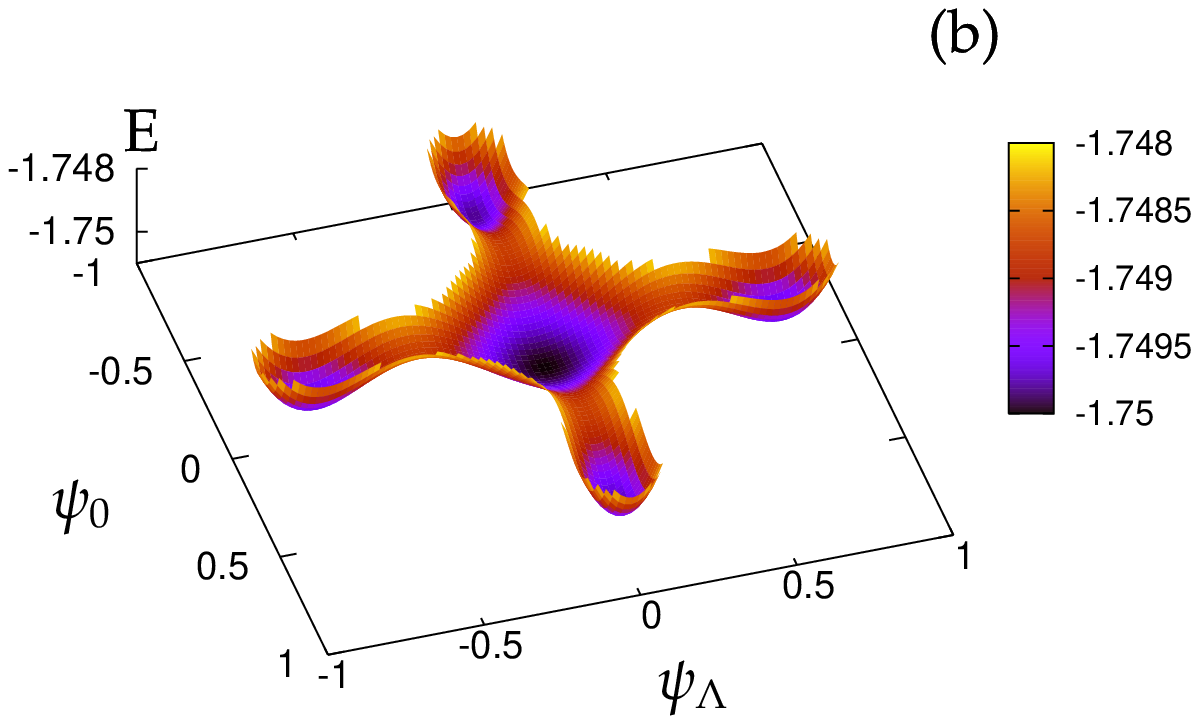,width=6cm,
clip}
\epsfig{figure=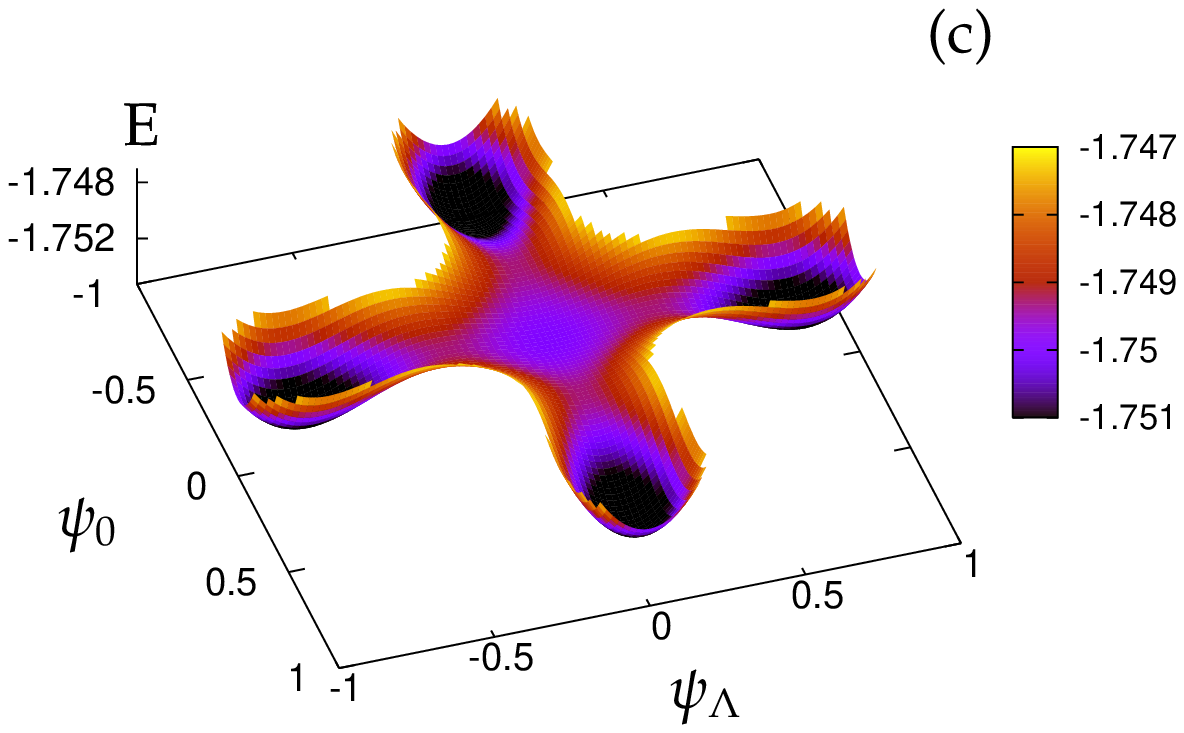,width=6cm,
clip}
\epsfig{figure=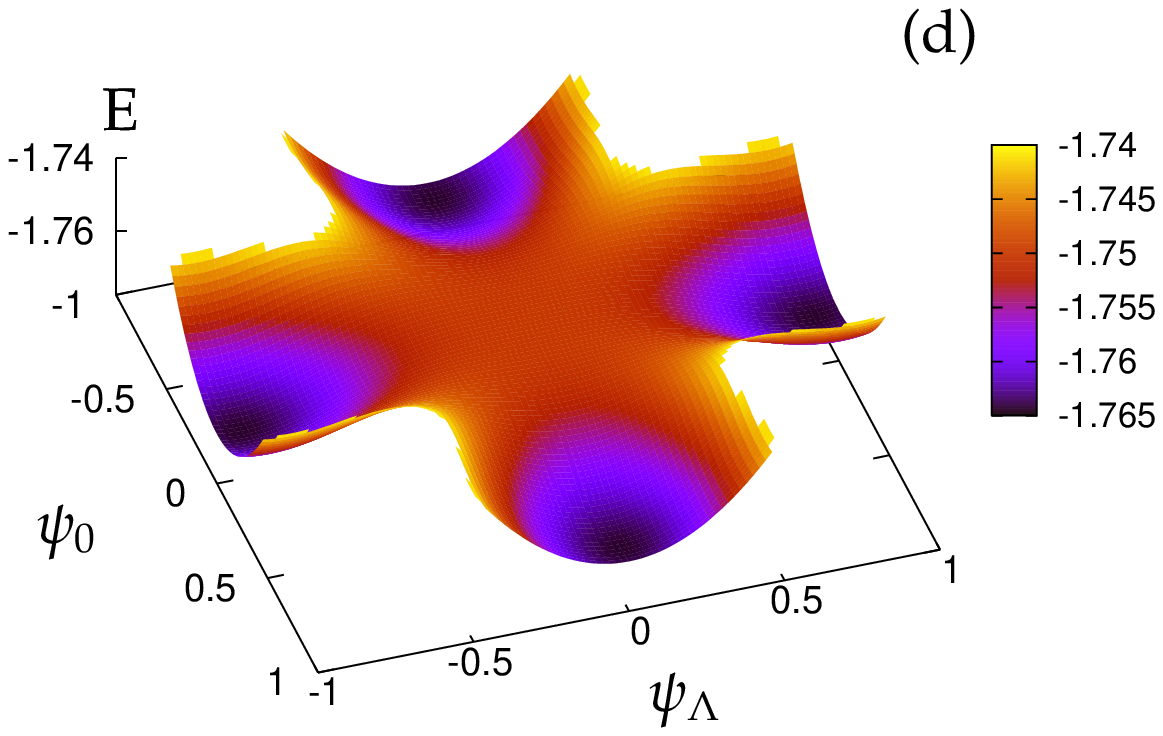,width=6cm,
clip}
\caption{(Color online) The ground state energy given by
Eq.~(\ref{mftham3}) in the anti-ferromagnetic case,
$U_2=0.1U_0$. (a) $U_0=33.3t$, the global minimum is at
$\psi_0=\psi_{\Lambda}=0$ and the system is in the MI phase. (b)
$U_0=12.5t$, the local minima are at non-zero $\psi_0$ or
$\psi_{\Lambda}$ indicating stable MI and metastable SF phases. (c)
$U_0=12.2t$, the global minima are at non-zero $\psi_0$ or
$\psi_{\Lambda}$ indicating stable SF and metastable MI phases. (d)
$U_0=11.11$, four degenerate global minima indicating the system is
in the SF phase.}
\label{antiferroEmin}
\end{figure}

\begin{figure}[ht]
\epsfig{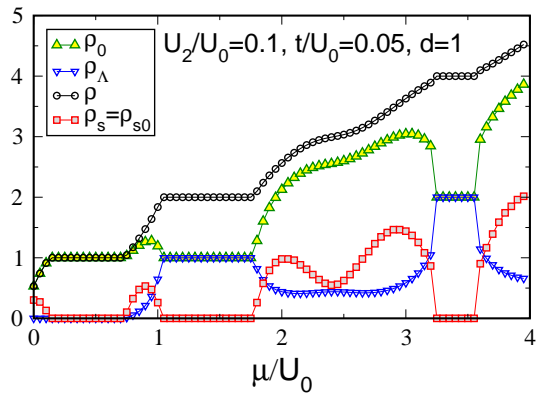}
\caption{(Color online) The densities of the two species, $\rho_0$ and
$\rho_{\Lambda}$, the total density, $\rho=\rho_0+\rho_{\Lambda}$, and
the corresponding superfluid densities as functions of $\mu/U_0$ in
the anti-ferromagnetic case. This cut is taken just outside the third
lobe (see Fig.~\ref{antiferroMFTphasediag}) so the compressibility,
$\kappa=\partial \rho/\partial \mu$, does not quite vanish. Note the
discontinuous transition into the fourth Mott lobe.
}
\label{antiferroMFTrhos}
\end{figure}

If the filling of the MI is even, then as the system approaches the tip
of the lobe, decreasing the interaction $U_0$, {\it local} minima
develop at non zero $\psi_0$ {\it or} $\psi_{\Lambda}$: With
Eq.~(\ref{mftham3}) the local minima do not form an axisymmetric
continuum as they do with Eq.~(\ref{mftham2}), Ref.~\cite{krutitsky04}.
As for the odd lobes just discussed, the minima lie either on the
$\psi_0$ axis or on the $\psi_{\Lambda}$ axis. This is illustrated in
Fig.~\ref{antiferroEmin} which shows, for $\rho=2$, the behaviour of the
energy minima as the interaction is reduced taking the system from the
MI to the SF phase. Figures \ref{antiferroEmin}(b,c) illustrate the
behaviour as the system is leaving the MI. First, local minima develop
with non-vanishing order parameter, indicating metastable SF, Fig.
\ref{antiferroEmin}(b). These local minima then become global, for
lower $U_0$, while the previously global minimum at the origin becomes
local indicating that the MI is now metastable, Fig.
\ref{antiferroEmin}(c). Such behaviour signals a first order SF-MI
transition for the even Mott phases. The nature of the transition is in
agreement with Ref.~\cite{krutitsky04} but not the nature of the SF
phase. According to our mean
field result, when the system is in the SF phase, superfludity is
carried entirely by one species or the other but not by both. This
result is in marked contrast with that of the MFT based on
Eq.~(\ref{mftham2}) and is similar to the MFT prediction for the SF
phase of the full spin-$1$ model \cite{pai08} in the antiferromagnetic
case.

\begin{figure}[ht]
\epsfig{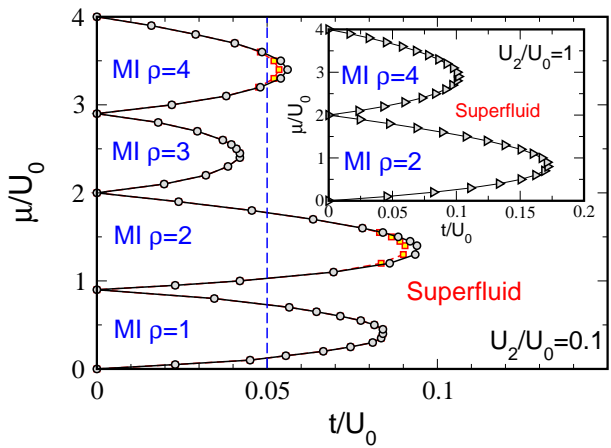}
\caption{(Color online) The mean field phase diagram given by
Eq.~\ref{mftham3} for $U_2/U_0=0.1$. The bases of the odd MI lobes are
reduced by $U_2/U_0$; the phase transitions near the tips of the even
lobes are first order. The regions between the squares and the
circles around the tips of the even lobes denote the regions where the
SF (MI) is metastable while MI (SF) is stable. Inset: The phase diagram
for $U_2/U_0=1$: all the odd MI lobes have vanished.}
\label{antiferroMFTphasediag}
\end{figure}

In Fig.~\ref{antiferroMFTrhos} we show the densities of the
two species, $\rho_0$ and $\rho_{\Lambda}$, the total density,
$\rho=\rho_0+\rho_{\Lambda}$, and the corresponding superfluid
densities as functions of $\mu/U_0$ in the anti-ferromagnetic case.
Several features are noteworthy. The first, second and fourth
incompressible MI plateaux are clearly visible but the third plateau
does not quite form. This is because the cut shown in Fig.
\ref{antiferroMFTrhos} passes just outside the third Mott lobe, see the
phase diagram Fig. \ref{antiferroMFTphasediag}. As in the ferromagnetic
case, the first Mott lobe is made of only one species, $\rho_0$ in this
case. In the even Mott lobes we have $\rho_0=\rho_{\Lambda}$ but nowhere
else. In addition, superfluidity is carried entirely by one species; in
this case $\rho_s=\rho_{s0}$ and $\rho_{s\Lambda}=0$ even though
$\rho_{\Lambda}\neq 0$. We also note that the transition into the
fourth plateau is discontinuous: We see clearly the discontinuous jumps of
$\rho_0$, $\rho_{\Lambda}$ and $\rho_s$ as this plateau is
approached. This is in agreement with the discussion of 
Fig.~\ref{antiferroEmin} and indicates that the system in 
Fig.~\ref{antiferroMFTrhos} enters the fourth Mott lobe near its tip. This
is confirmed in the full phase diagram Fig.~\ref{antiferroMFTphasediag}.

The phase diagram is obtained by calculating slices as in
Fig. \ref{antiferroMFTrhos} for different values of $\mu$ and $t/U_0$
at fixed $U_2/U_0$. Figure \ref{antiferroMFTphasediag} shows the
mean field phase diagram for the case $U_2=0.1U_0$ while the inset
shows it for $U_2=U_0$. The boundaries of the phases do not depend
much on whether one uses Eq.~(\ref{mftham2}) or (\ref{mftham3}) for the
mean field calculation. However the nature of the SF phase depends
crucially on which Hamiltonian is used as discussed above.
For the $U_2/U_0=0.1$ case, and more generally for small $U_2/U_0$, the
transition near the tip of even Mott lobes is first order, as discussed
in connection with Figs. \ref{antiferroMFTrhos}
and \ref{antiferroMFTphasediag}: The regions between the squares and the
circles around the tips of the even Mott lobes in Fig.
\ref{antiferroMFTphasediag} denote the regions where the SF (MI)
is metastable while MI (SF) is stable. The smaller the ratio $U_2/U_0$,
the wider the first order transition region. On the other hand,
when the ratio $U_2/U_0$ is large enough, there are no discontinuous
transitions; only continuous transitions remain. The extreme case of
$U_2=U_0$, where all odd lobes disappear, is shown in the inset to Fig.
\ref{antiferroMFTphasediag}.

\section{Quantum Monte Carlo Phase Diagram}

In this section we present the results of exact Quantum Monte Carlo
simulations for the system governed by the Hamiltonian Eq.~(\ref{ham2})
or (\ref{ham3}). As mentioned before, these two forms of the
Hamiltonian are identical for $U_2<0$ and are equivalent for $U_2>0$
in that they are related by a simple phase rotation and do, in fact,
give the same results in QMC simulations for all the quantities 
we looked at.

We use for our simulations the Stochastic Green Function (SGF)
algorithm \cite{SGF} with directed update \cite{directedSGF}. This
algorithm can be implemented both in the canonical and the grand
canonical ensembles \cite{fflobat}. In this work we used mostly the
canonical formulation where we relate the density to the chemical
potential using
\begin{equation}
\mu(N)=E(N+1)-E(N).
\label{mucan}
\end{equation}
$N$ is the total number of particles in a system with $L$ lattice
sites and $E=\langle H\rangle$ is
the total internal energy which is equal to the free energy in the
ground state. Another very important quantity to
characterize the phases is the superfluid density which is given by
\cite{roy}
\begin{equation}
 \rho_s = \frac{\langle W^2\rangle}{2dt\beta L^{d-2}},
\label{rhos}
\end{equation}
in the single species case where $W$ is the winding number of the
bosons. In the present case, there are two species of particles which
can be converted into each other. Consequently, the relevant winding
number is the total for the two species and the superfluid density is
then given by,
\begin{equation}
\rho_{\rm s} = \frac{\langle (W_0+W_{\lambda})^2\rangle}{2dt\beta
L^{d-2}}.
\label{rhosc}
\end{equation}

\subsection{Ferromagnetic case: $U_2<0$}

\begin{figure}[ht]
\epsfig{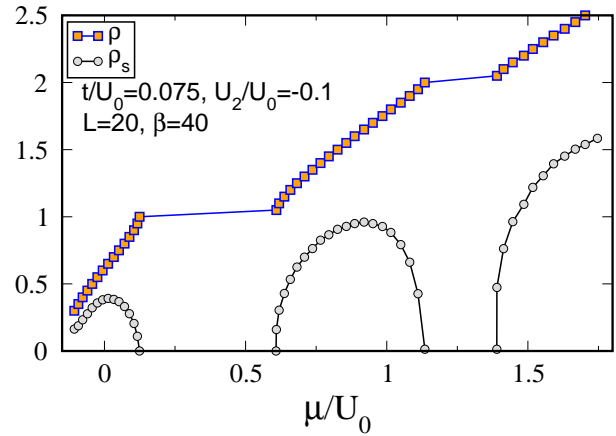}
\caption{(Color online) The total particle density, $\rho$ and the
superfluid density, $\rho_{\rm s}$ as
functions of the chemical potential, $\mu$.
The first two Mott plateaux are clearly visible.}
\label{ferrorhovsmuQMC}
\end{figure}

We start with ferromagnetic case, $U_2<0$. As we did with mean field,
we calculate the phase diagram in the $(\mu/U_0,t/U_0)$ plane at fixed
ratio $U_2/U_0$. We take $|U_2|/U_0=0.1$ because we want to use the same
value for the ferro- and the anti-ferromagnetic cases and, in the
latter case, MFT predicts first order transitions for the even Mott
lobes for this value. With $U_2/U_0=-0.1$, we study the SF-MI
transition as the chemical potential (or density) is varied by
calculating $\rho$ versus
$\mu$ slices for different fixed $t/U_0$. Using the canonical algorithm,
this is done by incrementing the number of bosons by one particle at a
time, doing the simulation, and then calculating the chemical
potential, $\mu$, using Eq.~(\ref{mucan}). Figure \ref{ferrorhovsmuQMC}
shows such a slice at $t/U_0=0.075$ and clearly exhibits the first two
incompressible Mott plateaux. Furthermore, we see that when the
system is compressible, it is also superfluid, $\rho_{\rm s}\neq 0$. In
the Mott plateaux, the superfluid density vanishes, $\rho_{\rm s}=0$.

\begin{figure}[ht]
\epsfig{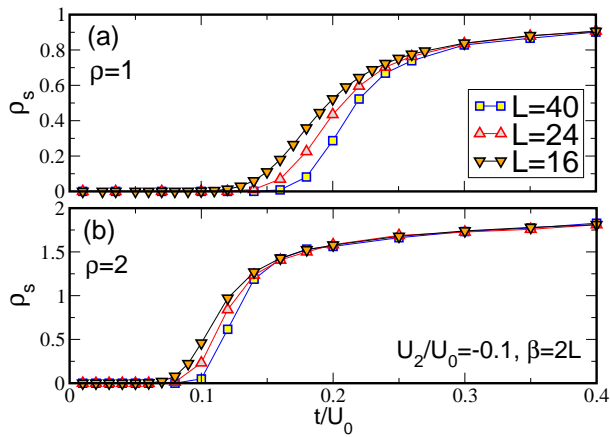}
\caption{(Color online) The superfluid
density, $\rho_{s}$, as a functions of $t/U_0$ in the
first (a) and second (b) Mott lobes. No evidence of a discontinuous
jump can be seen as the system size increases, the transitions are
continuous.}
\label{ferroMottsrhosQMC}
\end{figure}

Next, we study the SF-MI transition at fixed commensurate filling,
integer $\rho$, as the coupling, $t/U_0$ is changed. This exhibits the
behaviour of the transition at the tips of the Mott lobes.
In Fig. \ref{ferroMottsrhosQMC} we show the superfluid density,
$\rho_s$ versus $t/U_0$ for $\rho=1,\,2$. We see that in both cases,
$\rho=1$ (Fig.~\ref{ferroMottsrhosQMC}(a)) and $\rho=2$
(Fig.~\ref{ferroMottsrhosQMC}(b)), $\rho_s$ changes continuously as
predicted by MF.

\begin{figure}[ht]
\epsfig{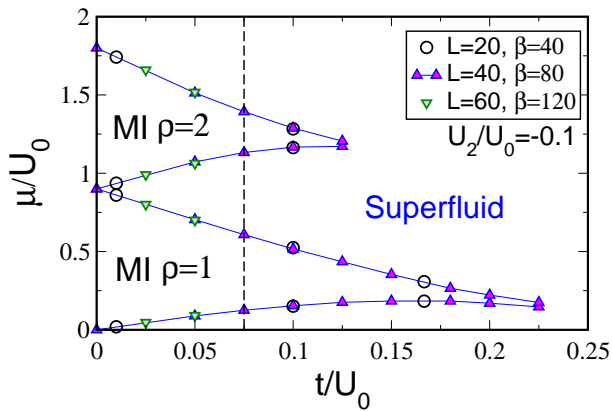}
\caption{(Color online) The QMC phase diagram of the ferromagnetic
system with $U_2/U_0=-0.1$. All Mott lobes 
shrink by $\delta\mu/U_0=0.1$ at their bases as in the mean field case,
Fig.~\ref{ferrophasediagMFT}. Note, however, that unlike the rounded
lobes in the mean field case, the lobes here end in cusps. This is
strong evidence that the gap vanishes exponentially, a behaviour
characteristic of the KT universality class of the two-dimensional $XY$
model. This phase diagram is very similar to that of the full spin-$1$
model in Ref.~\cite{batrouni}. The dashed vertical line indicates where
the cut in Fig.~\ref{ferrorhovsmuQMC} was taken.
}
\label{ferrophasediagQMC}
\end{figure}

The phase diagram is obtained by calculating $\rho$ versus $\mu$
scans for several $t/U_0$ values is shown in
Fig. \ref{ferrophasediagQMC}. System sizes $L=20,\, 40,\, 60$ were
used showing very little finite size effects for the gap. We remark
that, as expected, the Mott lobes shrink by $\delta\mu/U_0=0.1$
at their bases and will disappear completely for $|U_2|=U_0=1$. While
the QMC and MFT phase diagrams are similar qualitatively, the Mott
lobes given by the MFT are rounded while those given by exact QMC end
in cusps. This is consistent with a critical point at the tip in
the universality class of the two-dimensional $XY$ model which
is expected from general scaling arguments \cite{fisher} and
first observed in QMC in Ref.~\cite{batrouni90}.

\begin{figure}[ht]
\epsfig{figure=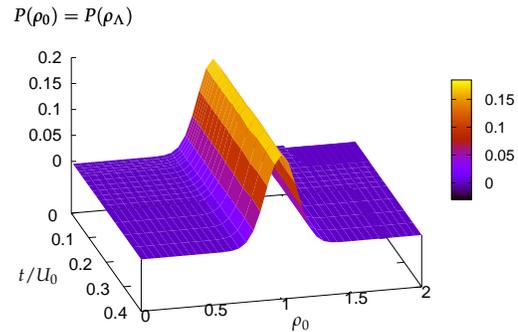,width=8cm,clip}
\caption{(Color online) The density distribution of the $0$ particles
as a function of $t/U_0$. The total density is $\rho=2$ and the values
of $t/U_0$ take the system from deep in the second Mott lobe to the
superfluid phase. The distributions of the two species, $0$ and
$\Lambda$, are identical and peak at a value of $\rho/2=1$. Compare
with Fig. \ref{antiferrohistogsrho2QMC}.}
\label{ferrohistogsQMC}
\end{figure}

Finally, we show in Fig. \ref{ferrohistogsQMC} the density
distribution for total density $\rho=2$ as $t/U_0$ is
increased taking the system from deep in the second Mott lobe into
the superfluid phase as in Fig.~\ref{ferroMottsrhosQMC}(b). We find that
the distributions for the $0$ and $\Lambda$ species are identical
and that the peak is at $\rho_0=\rho_{\Lambda}=1$ even though the
initial state for these simulations was chosen to
be $\rho_0=2,\,\rho_{\Lambda}=0$: In the ferromagnetic case, the density
distributions, $P(\rho_0)$ and $P(\rho_{\Lambda})$, are the same both in
the MI and SF phases. This means that the superfluid density is carried
by both particles. The same behavior was found for $\rho=1$.
This will be compared to the anti-ferromagnetic case
below.

\subsection{Anti-ferromagnetic case: $U_2>0$}

\begin{figure}[ht]
\epsfig{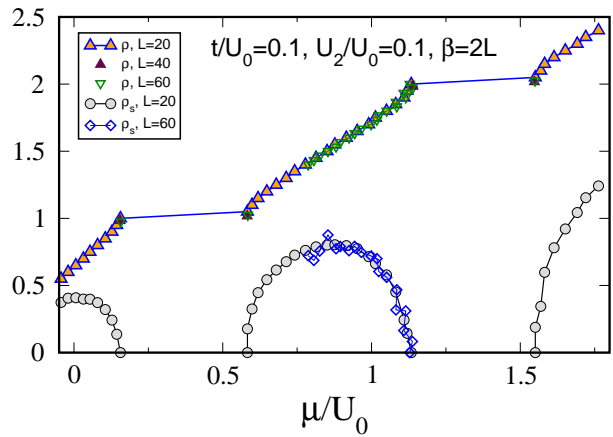}
\caption{(Color online) The total density, $\rho$ as a function of the
chemical potential, $\mu$, showing the first two incompressible Mott
phases. Also shown is the superfluid density $\rho_{\rm s}$ in the
compressible phases.
}
\label{antiferrorhovsmuQMC}
\end{figure}
\begin{figure}[ht]
\epsfig{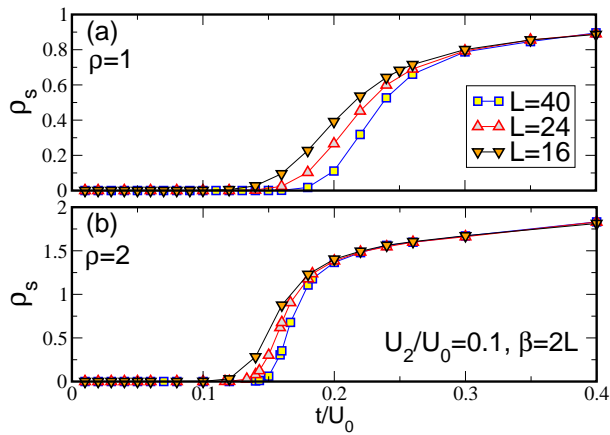}
\caption{(Color online) The superfluid density, $\rho_{\rm s}$, as
a function of $t/U_0$ in the first (a) and second (b) Mott lobes for
the antiferromagnetic case. No evidence of a discontinuous
jump can be seen as the system size increases. The SF-MI transitions
at the tips of the Mott lobes are continuous.
}
\label{antiferroMottsrhosQMC}
\end{figure}

The phase diagram in the anti-ferromagnetic case, $U_2>0$, is
determined in the same way as for $U_2<0$: We calculate slices of $\rho$
versus $\mu$ for various values $t/U_0$ keeping $U_2/U_0>0$ fixed. Such
a slice going through the first two Mott plateaux is shown in
Fig. \ref{antiferrorhovsmuQMC}. As in the ferromagnetic case, we see
that the compressible phase is superfluid, $\rho_{\rm s}\neq 0$.  Mean
field predicts that the SF-MI transition near the tip of the even
Mott lobes is first order. We found no evidence for that; $\rho_s$
vanishes continuously as a function of $\mu$ as the MI is approached.
In the canonical ensemble, as we have here, first order transition
are signalled by regions of negative compressibility
\cite{batrouniSS}, $\kappa=\partial \rho/\partial \mu<0$. We found that
$\kappa$ is always positive for all values of $t/U_0$ indicating that
the SF-MI transition is not first order.

\begin{figure}[ht]
\epsfig{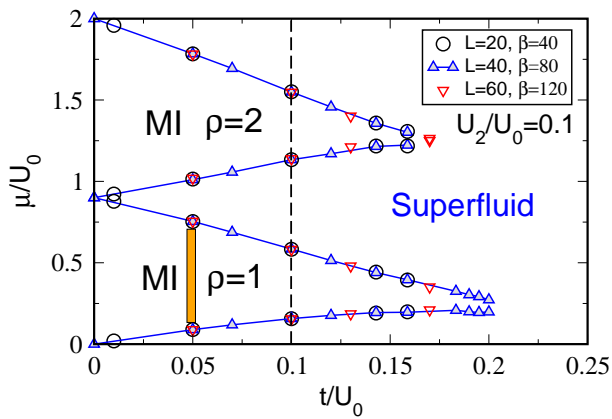}
\caption{(Color online) The phase diagram in the antiferromagnetic
case with data from three lattice sizes. The finite size effects are
small. Note that, as expected, the lower boundary of the base of the
second lobe shifts down by $U_2/U_0=0.1$ but the upper boundary does
not. When $U_2/U_0=1$, all odd order lobes disappear and only the even
order ones remain. The vertical dashed line indicates where the $\rho$
versus $\mu$ cut in Fig. \ref{antiferrorhovsmuQMC} was taken. The
vertical (orange) bar at $t/U_0=0.05$ in the first lobe indicates
where this Mott phase polarizes. See the discussion of Fig.
\ref{antiferrohistogsrho1QMC}.
}
\label{antiferrophasediagQMC}
\end{figure}

We confirm the nature of the SF-MI transition by studying the
behaviour of the superfluid density as a function of
$t/U_0$ at fixed total density $\rho=1,\,2$.
Figure \ref{antiferroMottsrhosQMC} shows $\rho_{\rm s}$ versus
$t/U_0$ for the first and second Mott plateaux. In both cases $\rho_s$
changes continuously as the system transitions from the MI to the SF
phase confirming that all SF-MI transitions in this model are
continuous.

\begin{figure}[ht]
\epsfig{figure=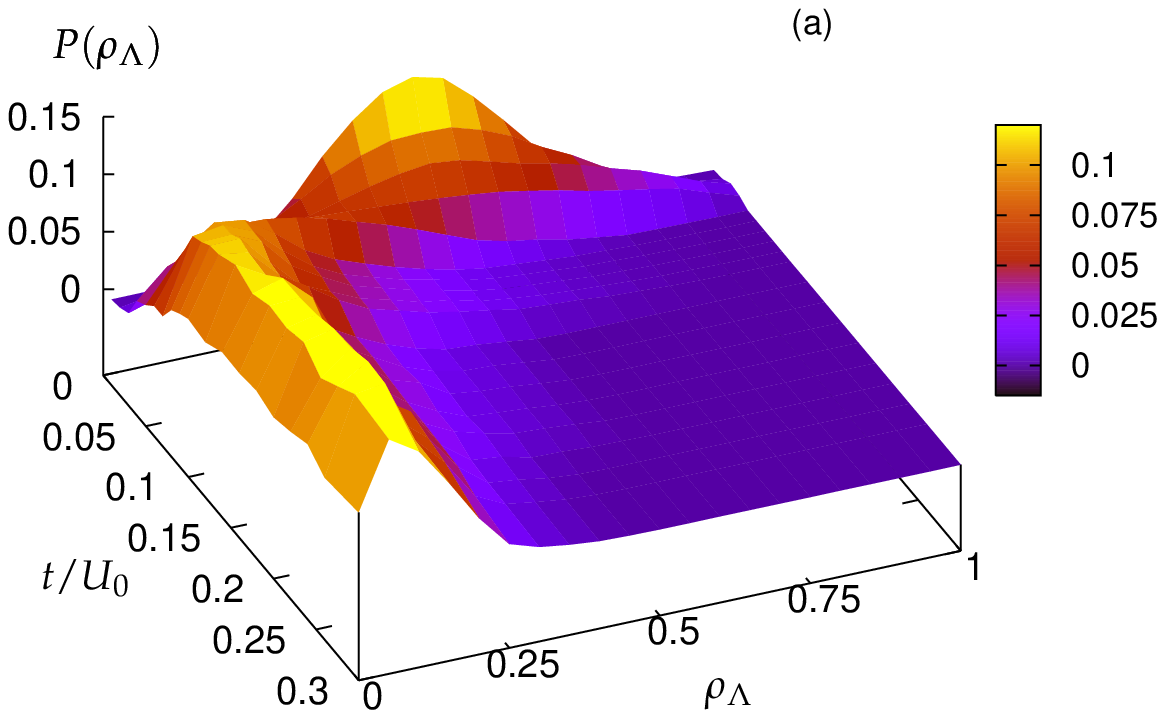,width=8cm,clip}
\\
\epsfig{figure=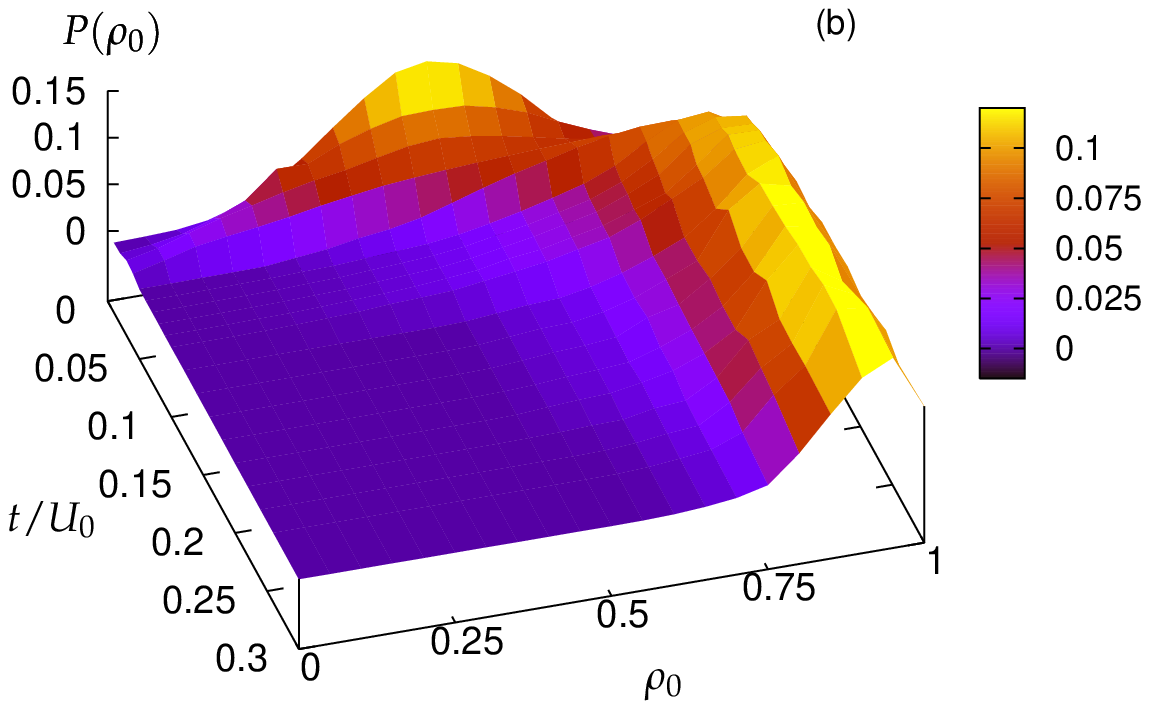,width=8cm,clip}
\caption{(Color online) The density distribution of the $\Lambda$ (a)
and $0$ (b) particles as functions of $t/U_0$. The system size is
$L=40$, $\beta=80$, and the total density is
$\rho=1$. The values of $t/U_0$ take the system from deep in the
first Mott lobe to the superfluid phase. Inside the Mott lobe, the
system is unpolarized, $P(\rho_0)=P (\rho_{\Lambda})$, for $t/U_0<
0.05$ and both distributions peak at $\rho/2=0.5$. At $t/U_0 \approx
0.05$, while the system is still in the first Mott lobe,  it
polarizes: The peak of one distribution approaches the total
density, $\rho=1$, while the other approaches $0$.  We verified that
the same behaviour is observed for different system sizes. This
transition in the Mott lobe is not predicted by MF.}
\label{antiferrohistogsrho1QMC}
\end{figure}

\begin{figure}[ht]
\epsfig{figure=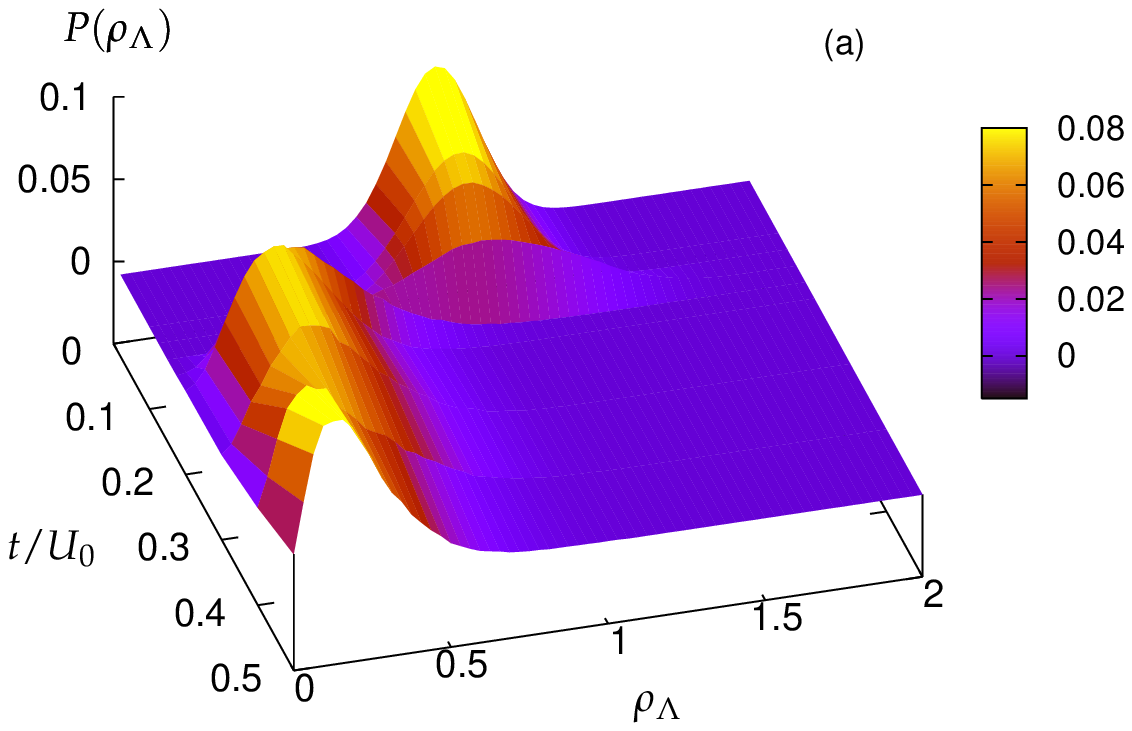,width=8cm,clip}
\\
\epsfig{figure=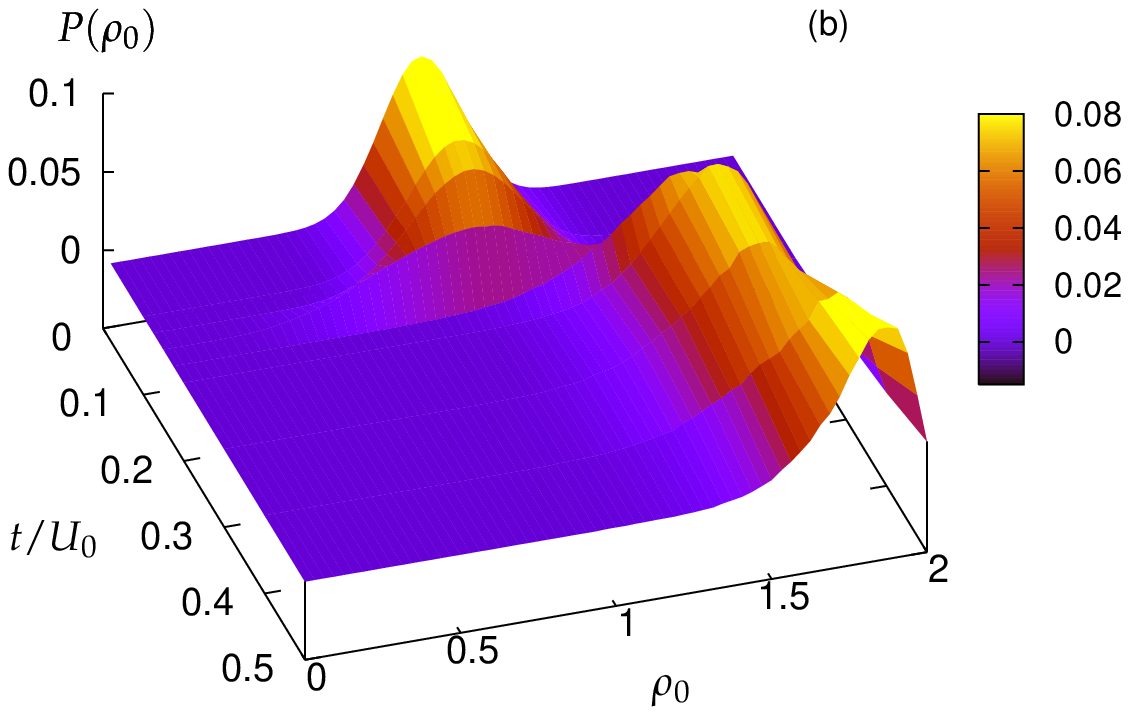,width=8cm,clip}
\caption{(Color online) The density distribution of the $\Lambda$ (a)
and $0$ (b) particles as functions of $t/U_0$. The system size is
$L=64$, $\beta=128$, and the total density is
$\rho=2$ and the values of $t/U_0$ take the system from deep in the
second Mott lobe to the superfluid phase. Inside the Mott lobe,
$P(\rho_0)=P (\rho_{\Lambda})$ and both distributions peak at
$\rho/2=1$. As soon as the system leaves the Mott phase and becomes
superfluid, it is polarized: The peak of one distribution approaches the
total density, $\rho=2$, while the other approaches $0$.  The
superfluid phase is polarized as predicted by the MFT: the
superfluidity is carried entirely by one species.
}
\label{antiferrohistogsrho2QMC}
\end{figure}

The phase diagram is obtained, as before, by calculating $\rho$ versus
$\mu$ for several values of $t/U_0$ and mapping out the Mott lobes. The
resulting phase diagram for the first two lobes is shown
in Fig.~\ref{antiferrophasediagQMC}.

Finally, we show in Fig. \ref{antiferrohistogsrho1QMC} that
the density distribution for total density $\rho=1$ as $t/U_0$ is
increased taking the system from deep in the first Mott lobe into
the superfluid phase. We see that for $t/U_0 <  0.05$, the system is
unpolarized, the two populations have identical distributions which
peak at $\rho_\Lambda=\rho_0 =0.5$. At $t/U_0\approx 0.05$, while the
system is still deep in the Mott lobe (see
Fig.~\ref{antiferrophasediagQMC}) the system polarizes: The population
of one species drops abruptly to near zero while that of the other
species increases to close to full filling. This polarization persists
through the SF-MI transition. This transition inside the first Mott
lobe is not predicted by mean field.

In Fig. \ref{antiferrohistogsrho2QMC} we show the density distributions
in the second Mott lobe. Here we find that the system polarizes only
when it transitions from the MI to the SF phase at $t/U_0 \approx
0.17$. Consequently, the SF phase in the antiferromagnetic phase is
always polarized: The system is mostly populated by a
single species and, consequently, superfluidity is carried by that
one particle type. This behaviour was predicted, qualitatively, by our
mean field calculation and is very different from the behavior in the
ferromagnetic case.

\section{Conclusions}

In this work, we performed a mean field calculation of the phase
diagram of the spin-$1/2$ bosonic Hubbard model, Eqs.~(\ref{ham2} and
\ref{ham3}). These models, which are derived from the full spin-$1$ model,
are equivalent when treated exactly but we
found that the mean field results depend on which Hamiltonian is used
in the antiferromagnetic ($U_2>0$) case but are identical in the
ferromagnetic case ($U_2<0$). Unlike a previous mean field result
\cite{krutitsky04} we found the superfluid phase is polarized when
$U_2>0$. We then used QMC to calculate the phase diagram for $U_2<0$
and $U_2>0$. For $U_2<0$, we found that the populations are balanced
both in the superfluid and in the incompressible phases. For $U_2>0$,
we found that the SF phase is always polarized with one species
dominating the population. Throughout the second Mott lobe (and
presumably in all even Mott lobes) the populations are balanced. On the
other hand, for $t/U_0 <0.05$ in the first Mott lobe, we found that
the populations are balanced but that the system polarizes inside the
Mott lobe at $t/U_0\approx 0.05$. This new transition inside the Mott
lobe was not predicted before.

\begin{acknowledgements}
This work was supported by: the CNRS-UC Davis EPOCAL LIA joint research
grant; by NSF grant OISE-0952300; an ARO Award W911NF0710576 with funds
from the DARPA OLE Program; We thank A. Brothers for helpful insight.
\end{acknowledgements}

\end{document}